\documentclass[10pt,conference]{IEEEtran}
\ifCLASSINFOpdf
\else
\fi

\usepackage{csquotes}
\usepackage{booktabs} 
\usepackage{framed}
\usepackage{graphicx}
\usepackage{amssymb}
\usepackage{tabularx}
\usepackage{graphicx}
\usepackage{url}
\usepackage{tabulary}
\usepackage{multirow}
\usepackage{enumitem}
\usepackage{hyperref}
\usepackage{hypcap}
\usepackage{lscape}
\usepackage[normalem]{ulem}
\useunder{\uline}{\ul}{}
\usepackage{longtable}
\usepackage{color}
\usepackage{hyperref}

\usepackage{booktabs}
\usepackage[flushleft]{threeparttable}
\usepackage{microtype}

\usepackage{microtype}
\usepackage{etoolbox}
\usepackage{makecell}
\usepackage[table]{xcolor}

\usepackage{times}

\usepackage{siunitx}  
\usepackage{textgreek}  
\usepackage{microtype}
\usepackage{textcomp}
\usepackage[caption=false]{subfig}

\usepackage{flushend}

\usepackage{etoolbox}
\newcommand{\ubold}{\fontseries{b}\selectfont} 
\robustify\ubold

\begin{document}
%
\title{The Daily Life of Software Engineers during the COVID-19 Pandemic}

\author{\IEEEauthorblockN{Daniel Russo}
\IEEEauthorblockA{Aalborg University\\ Department of Computer Science\\ Aalborg, Denmark\\
daniel.russo@cs.aau.dk}
\and
\IEEEauthorblockN{Paul H. P. Hanel}
\IEEEauthorblockA{University of Essex\\ Department of Psychology\\ Colchester, United Kindom\\
p.hanel@essex.ac.uk}
\and
\IEEEauthorblockN{Seraphina Altnickel}
\IEEEauthorblockA{mia raeumerei GmbH \\Berlin, Germany\\
}
\and
\IEEEauthorblockN{Niels van Berkel}
\IEEEauthorblockA{Aalborg University\\ Department of Computer Science\\ Aalborg, Denmark\\
nielsvanberkel@cs.aau.dk}
}


%


\maketitle

\begin{abstract}

Following the onset of the COVID-19 pandemic and subsequent lockdowns, software engineers' daily life was disrupted and abruptly forced into remote working from home. 
This change deeply impacted typical working routines, affecting both well-being and productivity.
Moreover, this pandemic will have long-lasting effects in the software industry, with several tech companies allowing their employees to work from home indefinitely if they wish to do so. 
Therefore, it is crucial to analyze and understand how a typical working day looks like when working from home and how individual activities affect software developers' well-being and productivity.
We performed a two-wave longitudinal study involving almost 200 globally carefully selected software professionals, inferring daily activities with perceived well-being, productivity, and other relevant psychological and social variables.
Results suggest that the time software engineers spent doing specific activities from home was similar when working in the office. 
However, we also found some significant mean differences. 
The amount of time developers spent on each activity was unrelated to their well-being, perceived productivity, and other variables.
We conclude that working remotely is not \textit{per se} a challenge for organizations or developers.
\end{abstract}


%
\IEEEpeerreviewmaketitle

\section{Introduction}
\label{sec:intro}


The SARS-CoV-2 (or COVID-19) pandemic disrupted abruptly software developers working routines in an unprecedented way.
Many software developers were asked to switch their typical office-based working habits to a new working from home (WFH) setting on short notice.
This has had a considerable negative impact on developers' well-being and productivity~\cite{Ralph2020pandemic}.
Nonetheless, research has also shown, using multiple-waives designs, that software engineers seem to adapt over time successfully, which has a positive effect on their well-being and productivity~\cite{ford2020tale,forsgren_2020,bao2020does,russo2020predictors}.
This is encouraging, as 89\% of professionals would like to work from home at least one day per month after the pandemic~\cite{Walton2020NZadaptation}.
Thus, there is a positive attitude towards remote working in the future.
For this reason, major IT companies (e.g., Twitter, Microsoft, AirBnB, Uber, Facebook) informed their employees that they could work from home indefinitely (e.g., Twitter) or extended the remote work policies providing specific support (e.g., AirBnB)~\cite{BusinessInsider2020}.


Remote work (or telework), \textit{per se} is not a new topic in software engineering.
With the rise of the internet in the late 90s, scholars started asking themselves about the challenges and opportunities of working from home~\cite{pounder1998homeworking}. 
Researchers investigated specific software development practices, such as process~\cite{guo2001special,deshpande2016remote} or communication~\cite{higa2000understanding}. 
Also, collaboration and characteristics of remote and asynchronous projects have been extensively studied by the Global Software Engineering community~\cite{herbsleb2007global,vsmite2010empirical}. 
Such studies typically focus on the interaction of software development teams co-located in different geographical areas.
However, the focus has been on software development teams working together on distributed projects.
So far, the research on working from home practices has been quite limited.
One reason is that managers are quite skeptical about remote working due to worries concerning employees' reduced focus, productivity, company culture, or team cohesiveness~\cite{buffer2020}. 
Nevertheless, the pandemic made many of us realize that some fears are unfounded (such as decreasing productivity) and that we have to face such challenges until a sufficient number of people have been vaccinated, a process that might take several years.
Hence, anecdotal evidence driving top managerial decisions due to the lack of specific research~\cite{mesaglio_2020} should be supplemented with scholarly evidence.
Thus, we formulate the following research questions:

\begin{quote}
    \textbf{Research Question$_1$}: How does the distribution of working activities of software engineers WFH during the pandemic compare to a pre-pandemic distribution of their working activities?
\end{quote}

\begin{quote}
    \textbf{Research Question$_2$}: Do well-being, productivity, and other psychological and social variables relate to developers' work activities while working remotely during the pandemic?
\end{quote}


Thus, in this paper, we explore how software development activities changed during the pandemic using the activity taxonomy of Meyer et al.~\cite{meyer2019today}, and whether specific activities contribute to software engineers' well-being and productivity. For example, there is countless anecdotal evidence that meetings are a waste of time~\cite{
Urban2019}. Does this imply that software developers' perceived productivity is lower when they have more meetings, and are more meetings also associated with lower well-being and more boredom? 
We further explore which activities are associated with well-being and stress. 
This research is also relevant because most previous research investigated predictors of well-being and stress in occupational settings~\cite{bhui2016perceptions,edwards2009value,mccalister2006hardiness} has not measured the specific activities that might have contributed to higher stress and lower levels of well-being. However, the type of activity someone is doing might contribute to higher stress levels beyond other factors identified by previous research, such as support by coworkers and supervisors~\cite{chyi2018prediction}. If we were to identify that specific activities are associated with higher or lower levels of stress or well-being, this would provide valuable information for future research investigating predictors of stress.  

Over a two-week period, we collected twice information regarding developers' activities and self-reported well-being and productivity measures to assess changes along with the lockdown.
We compared wave 1 with wave 2 to assess our test-retest reliability and stability of the data along with the pandemic.
In particular, we found that the time software engineers spent doing specific activities from home was overall comparable when working in the office.
Indeed, the working activities' rank remained almost the same, e.g., $coding > emails > code review > networking$. Nevertheless, we also reported some significant mean differences, such a lesser time dedicated to meetings and breaks and more specification and documentation. 
Furthermore, the amount of time people spent on each activity was not related to their well-being, perceived productivity, and other variables.

In the remainder of this paper, we describe the related work in Section~\ref{sec:related}, followed by a discussion about our research design in Section~\ref{sec:design}. 
The analysis and related results are described in Section~\ref{sec:analysis}.
Implications and recommendations for software engineers and organizations are then outlined in Section~\ref{sec:discussion}.
Finally, we conclude this study by outlying future research directions in Section~\ref{sec:conclusion}.

\section{Related Work}
\label{sec:related}

 Several large software companies, such as Stack Overflow or Red Hat, have embraced working from home by designing \textit{ad hoc} schemes~\cite{stackoverflow_2017, hat_2015}.
Organizations do so to increase their employees' job satisfaction and productivity while simultaneously reducing their operating expenses, such as office rent~\cite{felstead2017assessing, perez2002benefits}.
However, thus far, the software engineering literature did not primarily investigate working from home challenges, with a few exceptions.
To find previous work, we looked into peer-reviewed publications in Scopus. 
We identified six relevant papers.
Considering the vast but recent impact of COVID-19, we also selected non-peer-reviewed pre-prints on arXiv (three in total).
Table~\ref{tab:SLR} summarizes prior studies of remote working issues related to software engineers.

\begin{table*}[!htbp]
\centering
\caption{Overview of prior studies about software engineers working from home}
\label{tab:SLR}
\begin{tabular}{@{}m{2.6cm}m{7cm}m{7cm}@{}}
\toprule
\textbf{Study} & \textbf{Method} & \textbf{Findings} \\ \midrule
Bao et al. (2020)~\cite{bao2020does} & Field study. Mixed-methods study of 139 developers' during the COVID-19 lockdown at Baidu in China. Mining of 12 developers' activities over 138 days to investigate productivity while working from home. & Productivity depends on project characteristics (size, age, type, programming language). The average productivity does not change if working from home. To some developers, it is highly beneficial; to others, it is detrimental. \\ \addlinespace
Ford et al. (2020) ~\cite{ford2020tale} & Field study. Mixed-methods investigation of 3,634 Microsoft developers. Two surveys collected qualitative and quantitative insights about working from home conditions during the COVID-19 lockdown. & Quality of family life and time improved, although it might have led to a lack of focus, poor work-life boundaries, communications, and sync issues, developers adapt over time. \\ \addlinespace
Ralph et al. (2020)~\cite{Ralph2020pandemic} & Sample study. Large-scale cross-sectional study of 2,225 software developers globally working from home during the COVID-19 lockdown, surveying five variables. Data were analyzed using covariance-based structural equation modeling.   & Confirmation of a theoretical model. Professionals' well-being and productivity are suffering; well-being and productivity are strongly related to each other; women are disproportionately affected by this peculiar remote working setting. \\ \addlinespace
Russo et al. (2020)~\cite{russo2020predictors} & Sample study. Longitudinal study involving 192 software engineers living in countries with comparable COVID-19 lockdown measures, surveying 51 variables. Data were analyzed using correlations, multiple linear regressions, and covariance-based structural equation modeling to assess predictive-causal relations. & Well-being and productivity are related, professionals adapt to the condition over time, improving their well-being and productivity, introverts are disproportionally affected by the lockdown, no predictor variable was significantly able to causally explain the variance in well-being and productivity. \\ \addlinespace
Ford et al. (2019) ~\cite{ford2019remote} & Field study. Qualitative study interviewing three transgender software engineers to explore the interplay of gender identity and remote work. & Working from home enables the empowerment and identity disclosure of software professionals from marginalized communities. \\ \addlinespace
James \& Griffiths (2014)~\cite{james2014secure} & Experimental simulation. Within an existing project, relevant working from home problems have been identified and addressed by developing and validating a specific solution. & Development of a mobile execution
environment to support a secure and portable working from home setting. \\ \addlinespace
Guo (2001)~\cite{guo2001special} & Field study. Report of two qualitative surveys regarding software process improvement related to the distinctive characteristics of teleworking.  & Development of the \textit{Software Process Improvement approach for Teleworking Environment} (SPITE) model. Identification of 25 base practices to improve software processes when working from home. \\ \addlinespace
Higa et al. (2000)~\cite{higa2000understanding} & Field study. Mixed-methods study at Fujitsu with 44 software engineers to investigate how the use of E-mail influences telework. To test the hypotheses, three hierarchical regression models were used. &  An effective use of E-mails by remote workers leads to better work distribution and work productivity. \\ \addlinespace
Pounder (1998)~\cite{pounder1998homeworking} & Formal theory. Essay about security problems linked to telework. & This is the first paper that considers ``homeworking'' as a distinct working setting. It discusses the main security concerns and makes recommendations for organizations. \\ \bottomrule
\end{tabular}
\end{table*}

Our overview highlights how the subject matter arose with more extensive use of the internet (the late 90s), but it was simultaneously a relatively neglected topic until very recently.
Indeed, most of the paper has been published in 2019 onward and are dealing with the COVID-19 pandemic.
From a methodological perspective, most studies have been field studies involving a single company (e.g., Fujitsu, Baidu, Microsoft)~\cite{bao2020does,ford2020tale,higa2000understanding}.
Such real-world investigations aimed to understand the research phenomena by generating research hypotheses.
Two studies were conducted in a neutral setting on the opposite spectrum by asking participants a quantifiable judgment and analyzing such data through statistical techniques.
These two sample studies generalize their result on the entire software engineering population~\cite{Ralph2020pandemic,russo2020predictors}.

Content wise, half of the papers are concerned with specific topics related to working from home, such as security~\cite{pounder1998homeworking,james2014secure}, process~\cite{guo2001special}, work productivity~\cite{higa2000understanding}, and inclusion~\cite{ford2019remote}.
The other half mostly investigated well-being and productivity while working from home during the pandemic~\cite{ford2020tale,Ralph2020pandemic,russo2020predictors} and productivity-related to projects' characteristics~\cite{bao2020does}.

It is evident from the few related work that remote working in software engineering is an under-researched topic.
Possibly, one reason might be that businesses in the IT sector, allowing software professionals to work from home in a structured way, are relatively few~\cite{reynolds_2020}.
Most importantly, to this work, no one so far analyzed specific working activities while working from home and how this influences both the perceived productivity and well-being of software engineers.

\section{Research Design}
\label{sec:design}

To answer in a reliable and meaningful way our research questions, we employed a post-positivist epistemological stance, using a longitudinal design.
Carefully recruited software professionals were asked to complete the same survey twice, two weeks apart from the others.
Unique randomized IDs were assigned to participants to preserve their anonymity and track their individual participation across both waves. 


\subsection{Participants}
Before selecting our participants, we ran a power analysis to be sure to detect a small-to-medium effect size of $r$ = .20, using a power of .80 (for a two-sided test)\footnote{With $r$, we mean Pearson's r, which is a measurement of linear association between two variables; its values range between -1 and +1.}. 
As a result, we had to recruit at least 190 participants to obtain meaningful results (i.e., with enough statistical power).
Participants were selected from a broader set of 500 software engineers who have been carefully selected through a multistage process in a previous study by Russo \& Stol~\cite{russo2020gender}. 
From this pool, we only selected professionals working from home during the pandemic and live in countries with comparable lockdown measures.
Finally, we obtained a sample of 192 software engineers who completed the first survey ($M_{age}$~=~36.65 years, $SD$~=~10.77, range~=~19–63; 154 men, 38 women), and 184 of those who participated in the second wave. 
We provide demographic information on participants' gender, educational attainment, and location in Table~\ref{tab:demographics}.
We collected our data between 20--26 April 2020 (wave 1) and between 4--10 May 2020 (wave 2). 
To ensure high data quality \cite{palan2018prolific}, we recruited participants from the academic data collection platform Prolific Academic and compensated participants above the US minimum wage.
The survey was run using Qualtrics.

\begin{table}[]
\caption{Overview of sample's educational attainment, and location}
\centering
\begin{tabular}{@{}lll@{}}
\toprule
                             & \textit{\textbf{N}} & \textbf{\% of sample} \\ \midrule
Less than high school degree & 1                   & 0.5\%                 \\
High school graduate         & 9                   & 4.7\%                 \\
Some college but no degree   & 22                  & 11.5\%                \\
Bachelor's degree            & 97                  & 50.5\%                \\
Master's degree              & 52                  & 27.1\%                \\
Doctoral degree              & 10                  & 5.2\%                 \\
Other                        & 1                   & 0.5\%                 \\ \midrule
United Kingdom               & 61                  & 31.8\%                \\
United States                & 49                  & 25.5\%                \\
Ireland                      & 6                   & 3.1\%                 \\ 
Italy                        & 6                   & 3.1\%                 \\
Other                        & 70                  & 36.5\%                \\ \bottomrule
\end{tabular}
\label{tab:demographics}
\end{table}

\subsection{Measurements}

Several of the measurement values are derived from a related project. 
For a complete presentation of the used instruments, we directly refer to Russo et al.~\cite{russo2020predictors} and the Supplementary Materials. The longitudinal design also allowed us to compute test-retest reliabilities, $r_{it}$ (i.e., the stability of responses across two or more time-points), by correlating responses given by participants at time 1 with those at time 2 (we are using \textit{time} and \textit{wave} interchangeably), which provides additional information about a scale's reliability to the commonly used Cronbach's alpha~\cite{mcdonald2013test}. 
Coefficients close to 0 are undesirable since they indicate a low association between the two time-points, suggesting, among others, poor data quality.

\textbf{Activities.} We measured the same 15 activities that were measured by Meyer et al.~\cite{meyer2019today}. We did this because we believe they covered most activities and to have a pre-pandemic comparison group. We asked participants "During the past week, how much time did you spend on each task percentage-wise (\%)?" This was followed by the 15 activities (e.g., "Coding", "Email", "Bugfixing") which were rated on a slider-scale ranging from 0\% to 100\%. For the activities which might have been more ambiguous, a brief explanation was added in brackets such as "Helping (helping, managing or mentoring people)", "Networking (maintaining relationships)".

\textbf{Well-being}. We used the Satisfaction with Life Scale~\cite{diener1985satisfaction}. Our Cronbach's alpha\footnote{Cronbach's alpha is a measure of scale reliability. For exploratory research, using new measurement scales, values above .60 are desirable while for confirmatory research the threshold is above .70 (and below .95)~\cite{hair2013multivariate}.} values to measure internal consistency for both waves were the following $\alpha_{time 1}~=~.90$, $\alpha_{time 2}~=~.90$ ( $r_{it} = .72, p < .001$). 

\textbf{Productivity}. Measuring productivity in software engineering is a highly debated issue. Some scholars, for example, suggest making the measurement more objective by using function points~\cite{wagner2018systematic}. Ko has criticized this viewpoint as being detrimental in the long run~\cite{ko2019we}.
On the other hand, other researchers propose a self-reflection measure with developers' self-reporting their daily productivity~\cite{meyer2014software}.
In this work, we adopted a similar approach. 
We did not use a standard measure (e.g., such as Ralph et al.~\cite{Ralph2020pandemic} did).
Productivity was operationalized as a function of time spent working and efficiency per hour, compared to a typical week.
The reason for this choice is that we wanted to investigate the variance in productivity while working remotely as compared to being in the office ($r_{it} = .50, p < .001$). 

\textbf{Stress}. We used the Perceived Stress Scale~\cite{Cohen1988perceived}; $\alpha_1~=~.80$, $\alpha_2~=~.77$ ($r_{it} = .73, p < .001$).

\textbf{Boredom}. We used the Boredom Proneness Scale~\cite{farmer1986boredom,struk2017short}; $\alpha_1~=~.87$, $\alpha_2~=~.87$, ($r_{it} = .69, p < .001$).

\textbf{Autonomy, competence, and relatedness}. To measure the three needs of the self-determination theory~\cite{ryan2000self}, we used the psychological needs scale~\cite{sheldon2012balanced}. Need for autonomy's Cronbach's alpha level were: $\alpha_1~=~.72$, $\alpha_2~=~.76$ ($r_{it} = .76, p < .001$); for Competence: $\alpha_1~=~.77$, $\alpha_2~=~.65$ ($r_{it} = .76, p < .001$); and for Relatedness: $\alpha_1~=~.79$, $\alpha_2~=~.78$ ($r_{it} = .71, p < .001$).

\textbf{Quality and quantity of communication with colleagues and line managers}. We used a self-developed three items instrument ($\alpha_1~=~.88$, $\alpha_2~=~.92$; $r_{it} = .67, p < .001$).

\textbf{Daily Routines}. We developed a five items scale ($\alpha_1~=~.75$, $\alpha_2~=~.78$; $r_{it} = .73, p < .001$).

\textbf{Distractions at home}. We developed a two items scale ($\alpha_1~=~.64$, $\alpha_2~=~.63$; $r_{it} = .63, p < .001$).

\section{Analysis \& Results}
\label{sec:analysis}
\subsection{Changes in activities}
\begin{figure*}
  \centering
  \includegraphics[width=1\textwidth]{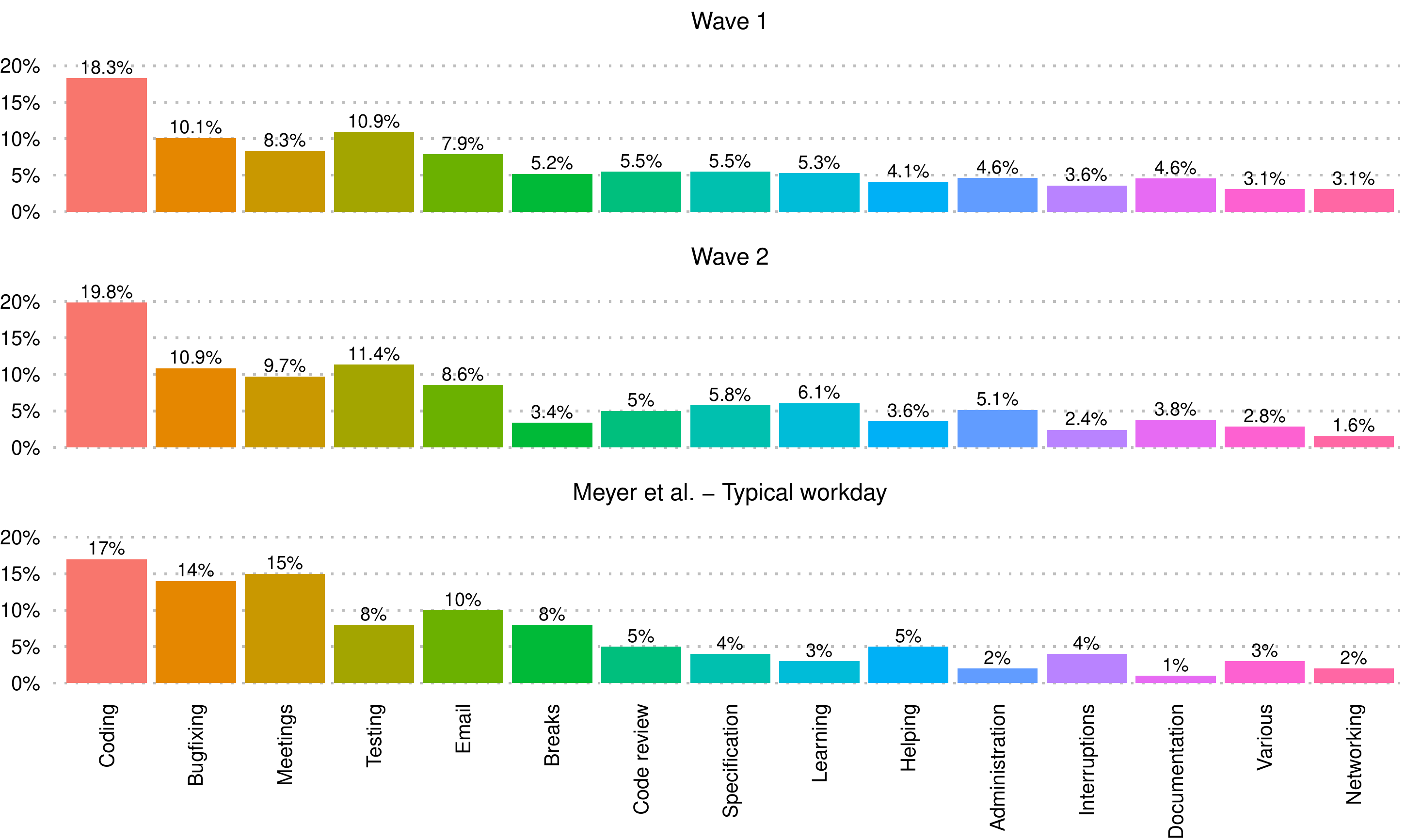}
  \caption{Distribution software engineering work activities during the two waves in our study, and a typical workday of software engineers as reported by Meyer et al.~\cite{meyer2019today}.}
  \label{fig:activities}
\end{figure*}

To test whether software developers' activities have changed during the pandemic, we first compared the time participants reported to have spent on each of the 15 activities with those reported by Meyer et al.~\cite{meyer2019today} (first research question). The results are displayed in Figure \ref{fig:activities}, as well as Tables \ref{tab:activities} and \ref{tab:comparison}. To test whether participants in our sample reported spending more or less of their time on certain activities than the software developers surveyed by Meyer et al.~\cite{meyer2019today}, we performed a series of one-sample $t$-tests. For example, we compared the percentages of participants in our sample at time 1 spend coding was significantly different from 15\%, which is the percentage reported by Meyer et al. (see Table \ref{tab:activities}, second column). We performed 15 (activities) $\times$ 2 (time points) = 30 $t$-tests (two-tailed, since we did not have directed hypotheses)\footnote{Because of the large number of comparisons, we adjusted the $\alpha$-threshold from .05 to .003 to reduce the risk of false-positive results. This means that we considered only $p$-values of $< .003$ as statistically significant. This is a standard procedure for studies that involve many variables to ensure reliable results, e.g., \cite{hanel2020well}. Note that changing the $\alpha$-threshold impacts the test statistic (e.g., $t-$value), as the test statistic and $p$-value are perfectly associated with any given sample size~\cite{hays_statistics_1994}. For example, for an $\alpha-$threshold of .003 and a sample size of 192 (time 1) or 184 (time 2), the critical $t$-values are 3.006 and 3.008. In other words, only if the $t-$value obtained from a $t-$test is larger than 3.006 (or 3.008), the $p$-value would be $<.003$, and we would consider the test result to be statistically significant. Note that a Bonferroni-correction would have resulted in an adjusted alpha-level of .05/30 $\approx$ .0017, which is overly conservative and does not consider that some of the variables are correlated (e.g., between time 1 and 2). Thus, the adjusted significance threshold of .003 seemed appropriate to us, neither overly conservative nor liberal.}. 
 \par

Software engineers in our sample reported on average to have spent less time bugfixing, in meetings, getting interrupted (only at time 2), helping (only at time 2), and taking breaks; but more time on testing, specification, writing documentation, networking (only at time 1), learning, and administrative tasks compared to the participants surveyed by Meyer et al. (Table \ref{tab:activities}). However, the differences between what our participants and those of Meyer et al. reported differed by only a few percent (see Figure \ref{fig:activities}). This visual inspection of the data is confirmed by correlation analysis. The 15 activities\footnote{For the correlations, the Degrees of Freedom are $N-2=13$ with N = 15 activities.} expressing percentages reported by Meyer et al. correlated with $r(13) = .84, p < .0001$ at time 1 and with $r(13) = .83, p = .0001$ at time 2. To obtain those correlations, we correlated the mean percentages reported in columns 2-4 of Table \ref{tab:activities} with each other. That is, we tested whether the average percentages spent on each activity reported by the participants in the Meyer et al. sample would align with those reported by the participants in our sample at wave 1 and 2. This suggests that while there are some deviations, the overall order of tasks remains stable. It further supports the quality of our data. If our participants had responded carelessly or even randomly, those two correlation coefficients would be around 0.

In a next step, we explored whether participants' activities changed over time. To do this, we performed a series of paired $t$-tests (Table \ref{tab:comparison}). The only statistically significant differences were observed for networking and taking breaks. At time 2, participants spent less time networking and taking breaks compared to time 1. Overall, the relative order of the activities remained very stable across time on the group level (i.e., when correlating the group averages of time 1 and 2), $r(13) = .99, p < .0001.$

\begin{table*}[]
\centering
\sisetup{
    group-digits=true,
    detect-weight=true,
    detect-shape=true,
    table-format=2.2,
    table-alignment = left
}
\caption{Comparisons of both waves with time spend on activities as reported by Meyer et al.~\cite{meyer2019today}}
\begin{tabular}{@{}lrrrS[table-format = -1.3]S[table-format = -1.3]S[table-format = <1.3]S[table-format = <1.3]@{}}
\toprule
\textbf{Activity}       & \textbf{Meyer et al.} & \textbf{M\textsubscript{t1}}    & \textbf{M\textsubscript{t2}}    & \textbf{$t$-value 1} & \textbf{$t$-value 2} & \textbf{$p$1}               & \textbf{$p$2}               \\ \midrule
Coding          & 17\%     & 18.11\% & 19.85\% & 0.901       & 1.89        & 0.369            & 0.060            \\
 Bugfixing      & 14\%     & 10.27\% & 10.85\% & -5.309      & -3.546      & <0.001 & <0.001 \\
 Meetings       & 15\%     & 8.45\%  & 9.74\%  & -9.951      & -6.628      & <0.001       & <0.001       \\
 Testing        & 8\%      & 10.96\% & 11.36\% & 3.413       & 3.321       & < 0.001 & 0.001  \\
 Email          & 10\%     & 7.93\%  & 8.59\%  & -3.686      & -1.584      & <0.001 & 0.115            \\
 Breaks         & 8\%      & 5.21\%  & 3.40\%  & -7.391      & -14.297     & <0.001 & <0.001 \\
Code review     & 5\%      & 5.44\%  & 5.01\%  & 0.878       & 0.019       & 0.381            & 0.985            \\
 Specification  & 3\%      & 5.49\%  & 5.76\%  & 4.653       & 4.048       & <0.001 & <0.001          \\
 Learning       & 3\%      & 5.30\%  & 6.07\%  & 4.242       & 3.377       & <0.001 & 0.001   \\
Helping         & 5\%      & 4.25\%  & 3.60\%  & -2.126      & -3.064      & 0.035            & 0.003   \\
 Administration & 2\%      & 4.70\%  & 5.15\%  & 4.575       & 4.279       & <0.001          & <0.001 \\
Interruptions   & 4\%      & 3.58\%  & 2.42\%  & -1.188      & -5.388      & 0.236            & <0.001       \\
 Documentation  & 1\%      & 4.69\%  & 3.77\%  & 5.178       & 5.073       & <0.001       & <0.001       \\
Various         & 3\%      & 3.17\%  & 2.84\%  & 0.592       & -0.346      & 0.554            & 0.729            \\
Networking      & 2\%      & 3.10\%  & 1.60\%  & 3.040        & -1.485      & 0.003   & 0.139            \\ \bottomrule
\end{tabular}
\\ \vspace*{3mm}
\footnotesize \textit{Note}. Activity percentages as per `typical workday' following Meyer et al.~\cite{meyer2019today}. M\textsubscript{t1}: mean at time 1 (see also Table \ref{tab:comparison}), $t$-value 1: $t$-value of one-sample $t$-test from time 1 vs value reported by Meyer et al., p1: $p$-value of one-sample $t$-test from time 1.\\
\label{tab:activities}
\end{table*}

\begin{table*}[]
\centering
\sisetup{
    group-digits=true,
    detect-weight=true,
    detect-shape=true,
    table-format=-1.3,
    table-alignment = left
}
\caption{Comparisons of activities between time 1 and time 2}
\begin{tabular}{@{}lrrrrSS[table-format = <1.3]SS[table-format = 2]S[table-format = 2]S[table-format = 2]@{}}
\toprule
 &  \multicolumn{2}{l}{\textbf{Time 1}} &  \multicolumn{2}{l}{\textbf{Time 2}} &   &   &   &   &   &   \\
 & \textbf{M} & \textbf{SD} & \textbf{M} & \textbf{SD} & \textbf{$t$} & \textbf{$p$} & \textbf{Cohen’s d} & \textbf{Higher} & \textbf{Smaller} & \textbf{Equal} \\ \midrule
Coding          &  18.11\% &  16.973\%    &  19.85\%  &  20.444\% &  -1.502 &  0.135 &  -0.108 &  94 &  74 &  15 \\
Bugfixing       &  10.27\% &  9.722\%     &  10.85\%  &  12.038\% &  -0.422 &  0.673 &  -0.037 &  68 &  86 &  29 \\
Meetings        &  8.45\%  &  9.103\%     &  9.74\%   &  10.767\% &  -2.418 &  0.017 &  -0.153 &  78 &  69 &  36 \\
Testing         &  10.96\% &  11.970\%    &  11.36\%  &  13.720\% &  -0.205 &  0.838 &  -0.014 &  74 &  85 &  24 \\
Email           &  7.93\%  &  7.776\%     &  8.59\%   &  12.103\% &  -0.705 &  0.482 &  -0.063 &  72 &  85 &  27 \\
 Breaks         &  5.21\%  &  5.208\%     &  3.40\%   &  4.362\% &  4.705 &  <0.001 &  0.367 &  47 &  102 &  33 \\
Code review     &  5.44\%  &  6.967\%     &  5.01\%   &  7.924\% &  0.385 &  0.700 &  0.035 &  56 &  76 &  50 \\
Specification   &  5.49\%  &  7.407\%     &  5.76\%   &  9.251\% &  -0.194 &  0.847 &  -0.016 &  54 &  68 &  61 \\
Learning        &  5.30\%  &  7.459\%     &  6.07\%   &  12.313\% &  -1.046 &  0.297 &  -0.089 &  51 &  76 &  55 \\
Helping         &  4.25\%  &  4.872\%     &  3.60\%   &  6.184\% &  1.664 &  0.098 &  0.128 &  46 &  81 &  57 \\
Administration  &  4.70\%  &  8.143\%     &  5.15\%   &  9.976\% &  -0.706 &  0.481 &  -0.051 &  55 &  80 &  47 \\
Interruptions   &  3.58\%  &  4.811\%     &  2.42\%   &  3.981\% &  2.814 &  0.005 &  0.263 &  39 &  79 &  62 \\
Documentation   &  4.69\%  &  9.841\%     &  3.77\%   &  7.411\% &  1.256 &  0.211 &  0.116 &  50 &  71 &  62 \\
Various         &  3.17\%  &  3.974\%     &  2.84\%   &  6.384\% &  0.590 &  0.556 &  0.051 &  49 &  78 &  56 \\ 
 Networking     &  3.10\%  &  4.977\%     &  1.60\%   &  3.674\% &  4.334 &  <0.001 &  0.350 &  31 &  77 &  74 \\ \bottomrule
\end{tabular}
\\ \vspace*{3mm}
\footnotesize \textit{Note}. t: $t$-value of a dependent sample $t$-test; Cohen's d: standardized mean difference; Higher: Participants who scored higher on an activity at time 2 compared to time 1; Lower: Participants who scored lower at time 2; Equal: Number of participants whose score has not changed.
\label{tab:comparison}
\end{table*}

\subsection{Correlates of activities}
To test the second research question, we correlated the time participants spent on each activity with the selected variables. This was possible because the activities were mostly uncorrelated in both time points on an individual level. We report Pearson's correlation ($r$) in our tables since most of the distributions were normally distributed. However, for the sake of completeness, we also ran a non-parametric Spearman's rank correlations test (reported in the Supplementary Material), which provided us with very similar results, suggesting the robustness of our results.
In total, we computed at both time points 13 (well-being related variables and productivity) $\times$ 15 (activities) = 195 correlations. Given a large number of comparisons, we changed our significance threshold from $\alpha = .05$ to .0005. Note again that a Bonferroni-correction would have resulted in an adjusted alpha-level of .00017, which is overly conservative and does not consider that some of the variables are correlated (e.g., distractions and stress). Thus, the adjusted significance threshold of .0005 seemed appropriate to us, neither overly conservative nor liberal. This new threshold implies that only correlation coefficients of $r \geq .25$ are significant. This is because the $p$-value of $r = .25$ is just below the .0005 threshold for our sample size of 192, $p \approx .00047$.  \par

\begin{table*}[]
\centering
\sisetup{
    group-digits=true,
    detect-weight=true,
    detect-shape=true,
    table-format=-1.2,
    table-alignment = left
}
\caption{Correlations between activities and variables at Time 1}
\resizebox{\textwidth}{!}{%
\begin{tabular}{@{}lSSSSSSSSSS@{}}
\toprule
 &
  {Well being} &
  {Productivity} &
  {Stress} &
  {Boredom} &
  {Relatedness} &
  {Competence} &
  {Autonomy} &
  {Communication} &
  {Daily routines} &
  {Distractions} \\ \midrule
Coding         & 0.09  & -0.02 & -0.20  & -0.04 & 0.15  & 0.13  & 0.18  & 0.08  & 0.11  & -0.10  \\
Bugfixing      & 0.03  & 0.09  & -0.11 & -0.14 & -0.04 & 0.03  & 0.02  & 0.08  & 0.03  & 0     \\
Meetings       & -0.08 & 0.13  & 0.14  & 0.01  & -0.11 & -0.02 & -0.25 & 0     & -0.07 & -0.05 \\
Testing        & -0.02 & -0.01 & -0.04 & -0.06 & 0.13  & 0.06  & -0.02 & -0.06 & 0.15  & -0.02 \\
Email          & -0.08 & 0.12  & 0.04  & -0.05 & -0.07 & -0.05 & -0.05 & 0.06  & 0     & -0.02 \\
Breaks         & 0     & \ubold -0.30  & 0.14  & 0.17  & -0.07 & -0.18 & 0.01  & -0.10  & -0.07 & 0.13  \\
Code review    & 0.13  & 0.08  & -0.11 & -0.03 & 0.04  & 0.17  & 0.06  & 0.12  & 0.11  & -0.11 \\
Specification  & 0     & 0.09  & 0.05  & 0.02  & -0.03 & -0.11 & -0.12 & -0.01 & -0.05 & 0.11  \\
Learning       & -0.07 & -0.07 & 0.13  & 0.12  & -0.05 & -0.11 & 0.05  & 0.06  & -0.15 & 0.11  \\
Helping        & 0.07  & 0.10   & -0.08 & -0.12 & 0     & 0.12  & -0.02 & 0.03  & 0     & -0.14 \\
Administration & 0.03  & -0.11 & -0.02 & -0.01 & 0.02  & 0.01  & 0.06  & -0.14 & -0.05 & 0.07  \\
Interruptions  & -0.21 & 0     & 0.20   & 0.07  & -0.27 & -0.21 & -0.20  & -0.08 & -0.21 & 0.12  \\
Documentation  & -0.03 & -0.07 & 0.09  & 0.05  & -0.03 & -0.05 & -0.01 & -0.07 & -0.01 & 0.03  \\
Various        & -0.08 & -0.11 & 0.07  & 0.02  & -0.03 & -0.08 & -0.04 & -0.06 & -0.11 & 0.13  \\
Networking     & 0.06  & 0.08  & 0.10   & 0.15  & 0.03  & 0.07  & 0.07  & -0.02 & -0.03 & -0.05 \\ \bottomrule
\end{tabular}
}
\label{tab:activities1}
\end{table*}

\begin{table*}[]
\centering
\sisetup{
    group-digits=true,
    detect-weight=true,
    detect-shape=true,
    table-format=-1.2,
    table-alignment = left
}
\caption{Correlations between activities and variables at Time 2}
\resizebox{\textwidth}{!}{%
\begin{tabular}{@{}lSSSSSSSSSS@{}}
\toprule
 &
  {Well being} &
  {Productivity} &
  {Stress} &
  {Boredom} &
  {Relatedness} &
  {Competence} &
  {Autonomy} &
  {Communication} &
  {Daily routines} &
  {Distractions} \\ \midrule
Coding         & 0.11  & 0.02  & -0.07 & 0.01  & 0.14  & 0.08  & 0.19  & 0.12  & 0.13  & 0     \\
Bugfixing      & 0.07  & 0.15  & -0.07 & -0.02 & -0.01 & 0.01  & 0.06  & 0.05  & 0.12  & -0.03 \\
Meetings       & -0.09 & 0     & 0.02  & -0.02 & -0.03 & -0.01 & -0.17 & 0.01  & -0.03 & -0.02 \\
Testing        & 0.03  & 0.07  & 0.04  & -0.08 & 0.08  & -0.02 & -0.02 & 0.01  & 0     & -0.09 \\
Email          & -0.13 & -0.06 & 0.01  & 0.03  & -0.09 & -0.05 & 0.01  & -0.21 & -0.10  & 0.05  \\
Breaks         & -0.11 & -0.16 & 0.03  & 0.16  & -0.09 & -0.15 & -0.01 & -0.08 & -0.02 & 0.07  \\
Code review    & -0.02 & -0.05 & 0.07  & 0.11  & -0.01 & -0.05 & -0.14 & -0.09 & -0.07 & 0.03  \\
Specification  & 0     & 0.09  & 0.03  & 0.10   & -0.12 & -0.01 & -0.10  & 0.18  & -0.02 & 0.01  \\
Learning       & 0.03  & -0.21 & 0.06  & 0.03  & 0.06  & -0.01 & 0.06  & -0.01 & -0.06 & 0.17  \\
Helping        & 0.01  & 0.03  & -0.11 & -0.19 & 0.16  & 0.13  & 0.02  & 0.12  & 0.01  & -0.13 \\
Administration & -0.09 & -0.05 & 0.09  & -0.02 & -0.10  & -0.04 & -0.11 & -0.18 & -0.10  & 0.03  \\
Interruptions  & -0.08 & 0.04  & 0.05  & -0.05 & -0.04 & -0.02 & -0.06 & 0     & 0.03  & -0.05 \\
Documentation  & 0.01  & 0.13  & -0.03 & -0.04 & -0.13 & 0.02  & -0.05 & -0.05 & -0.07 & -0.01 \\
Various        & 0.03  & -0.03 & -0.01 & 0.09  & -0.11 & -0.06 & 0.03  & -0.03 & 0.02  & -0.05 \\
Networking     & 0.04  & -0.07 & -0.13 & -0.13 & 0.07  & 0.08  & 0.05  & 0.10   & 0.06  & -0.11 \\ \bottomrule
\end{tabular}
}
\label{tab:activities2}
\end{table*}

The correlation coefficients are presented in Table \ref{tab:activities1} and Table \ref{tab:activities2}. 
This analysis did not show substantially significant results.
At time 1, only productivity was negatively correlated with time spent on breaks, $r = -.30, p = .00002$, which can be more considered to validate further our productivity measure rather than a meaningful finding itself. At time 2, none of the correlations was significant at $\alpha = $.0005. The correlation between productivity and time spent on breaks was again negative but did not reach statistical significance, $r = -.16, p = .03$.
Overall, we conclude that work activities carried out at home are not related to the identified variables.

\section{Discussion}
\label{sec:discussion}

\subsection{Implications for Research and Practice}

Working from home (WFH) has thus far been poorly considered by the software engineering literature as a contingent research topic.
Our investigation addresses the need to provide scholarly evidence concerning how working from home during the COVID-19 pandemic affected software developers' working activities.
Further, a deeper understanding of the emergent phenomenon's professional effects for a large number of software professionals working remotely provides relevant insights for both research and practice. 
To this end, this study makes several contributions.

First, we ran a longitudinal analysis of 192 carefully selected software professionals during the COVID-19 lockdown.
We assessed developers' working activities and their perceived well-being, productivity, and other relevant psychological and social variables.
Our data quality was assured by the test-retest reliability of each variable measuring at least $.50$, and Cronbach's alpha above $.60$.
Second, we compared the time spent on typical office-based working activities with the same activities while working from home.
Using the taxonomy and previously collected data of Meyer et al.~\cite{meyer2019today}, we ran 30 one-sample $t$-tests to assess significant differences.
Although we reported some differences, they are relatively small, which indicates that the time spent on different activities is almost identical in both working environments.
Third, we analyzed whether the time spent on each working activity changed during the pandemic.
After performing 15 paired $t$-tests, we conclude that developers spent their time consistently while working from home.
Fourth, we investigated the influence each identified variable had on the working activities, and if such an outcome is stable over time.
To do so, we ran 195 correlation analysis.
Our results suggest that the measured variables and activities are not correlated.
Fifth, we outline practical, evidence-based implications, as summarized in Table~\ref{tab:Findings}.

On the whole, we did not register significant changes regarding developers' work distribution. 
Further, we highlight that Meyer et al. 's sample refers only to one software company, whereas we surveyed developers across many companies, globally distributed.
Thus, some deviations were expected.
Nevertheless, we still report an overall consistency between our WFH data and Meyers et al.'s analysis of a typical office day at Microsoft.
Our results, therefore, show that working from home does not affect how software engineers dedicate their time to specific tasks.  
On a precautionary note, the reader should be aware that we did not monitor developers' effectiveness by executing every task while working remotely.
We opted for this choice to be consistent with Meyer et al., and because we collected data from a global sample of software professionals working in 190+ different organizations, making the development of objectively comparable measurements near impossible.
Still, we report some differences with the data collected by Meyer et al., although the difference is of only some percentage points. 
Most notably, software engineers spend less time in bugfixing, meetings, and breaks.
Also, they report fewer interruptions and less time on e-mail writing when working from home (although this is the case in only one of the two waves).
Contrary, they spend more time on specifications, testing, administration, documentation, and learning.
From these results, we notice that meetings are significantly reduced while working remotely, meaning that they are, on average shorter and more time-efficient than in the office. 
Also, participants invested in improving their skillset, spending more time on learning.
Similarly, developers seem to be more focused on their tasks, considering fewer reported breaks and interruptions.
However, this does not mean that they are not linked to their organization or their colleagues, since the time spent on networking remained the same.
This cautiously suggests that WFH might be more beneficial for both developers and organizations than working in the office, or at least for some group of professionals \cite{ford2019remote}.
Another limitation is that we only measured the time participants spent on each of the 15 activities using percentages rather than absolute time (e.g., in minutes). This implies that the comparisons we made between wave 1 and 2 as well as the sample from Meyer et al.~\cite{meyer2019today} are in relative terms but not absolute terms. For example, while participants in our sample reported to have spent only 8.45\% (wave 1) and 9.74\% (wave 2) of their time in meetings, and Meyer et al.'s participants reported to had spent 15\%, participants in our sample might have worked more and spent in absolute terms the same amount of time in meetings.  

During the pandemic, we did not register any significant change in the work activities, with only two exceptions: at the first wave, developers spent more time for breaks and networking than the second wave. 
Nevertheless, we report a correlation close to 1 of the group averages, suggesting a very high consistency of the activity distribution along with the pandemic.
The reason software engineers spent less time on breaks and networking during the second measurement point might indicate that they became more used to their new WFH condition.
Accordingly, professionals learned to spend their working time more efficiently.
Unfortunately, we did not collect any additional data that might support this point.
However, similar conclusions are supported by the literature~\cite{ford2020tale,russo2020predictors}.

Finally, we did not find any significant results from our extensive correlation analysis between working activities and potentially relevant variables (with one exception). 
This is a generally positive finding, as it shows that important psychological and social variables have no direct influence on developers' working tasks while working remotely.

The only significant relation was productivity, which correlated negatively with breaks in wave 1.
Despite being intuitive, we are very cautious about concluding that developers should take fewer breaks to be more productive since such a relation was not significant at wave 2 (although still negative).
Regarding the other activities, we conclude that the time spend on each task does not affect productivity.
Similar considerations can be made with well-being. 
We did not register any significant effect on how the amount of time dedicated to development activities impacts software engineers' well-being working from home.
We consider this evidence supportive of an extensive working from home setting since developers' well-being and productivity are not related to individual working tasks, meaning that organizations can plan working schedules as remote workers would still work from the office.

Furthermore, the stress in particular, which is an indicator of burnout~\cite{chyi2018prediction}, seems not to relate with any particular activity, meaning that none of the performed work tasks are \textit{per se} associated with stress. 
This is reassuring as it suggests that no activity causes stress, burnout, or lower well-being levels.
We can draw similar conclusions for the other identified variables.
None of the working activities were related to boredom.
This means that while working remotely, developers did not felt any specific task as dull and demotivating. 
Previous studies showed that during the pandemic, it is essential to have daily routines to improve personal well-being \cite{russo2020predictors}.
However, when it comes to individual activities, routines seem not to play a significant role.
Regardless of how software engineers organize their day, this does not affect the amount of time they dedicate to one activity or another.
Likewise, possible distractions that might happen while working from home (e.g., children at home) does not influence the time spent on work activities.
This is also a very relevant result, as the literature suggests that distractions play a significant role when working remotely \cite{Ralph2020pandemic}.
Although we do not contradict the conclusion of Ralph et al., we do report that distractions \textit{per se} are not related to developers' working tasks. 
Self determination theory measures innate psychological needs~\cite{ryan2000self}, and its three dimensions need for autonomy, competence, and relatedness are associated with work motivation in general~\cite{gagne2005self}. 
To the best of our knowledge, this study is the first in our community to assess whether specific activities are correlated with autonomy, competence, and relatedness. 
We found overall that psychological needs were unrelated to people's specific activities. 
This means that developers were not generally unmotivated (or motivated) with the time spent on the working tasks performed remotely. 
While working remotely, quality of communication can be challenging, as face-to-face communication has to pass through a medium (e.g., MS Teams, Zoom).
Not being directly connected to the organizations can, therefore, become a big issue for remote workers.
For example, research suggests that lower support from coworkers and supervisors~\cite{mccalister2006hardiness}, perceiving the values of one's organization to be different to one's values~\cite{edwards2009value}, and unfair treatment and lack of appreciation~\cite{bhui2016perceptions} are putting the mental health of remote workers at risk. 
Interestingly, our results suggest that the quality of communication does not relate to individual working activities.
This can also be considered a positive finding, as the time spent by software engineers for each task is not detrimental to the relations with their organization.

Prior research has mostly ignored whether activity type plays a role in professionals' psychological and social factors. 
Typically, scholars only measured whether people are, e.g., stressed overall, as opposed to stressed by specific activities~\cite{bhui2016perceptions, edwards2009value, mccalister2006hardiness}. 
Our research suggests that the type of activity is not a confounding variable, which increases our trust in prior research, which has typically looked at subjective work experience in general rather than actual activities. 
To conclude, our findings imply that software engineers' psychological and social factors do not matter on \textit{what} work activity they are performing, but rather \textit{how} it is done.

\begin{table*}[!ht]
\centering
\caption{Summary of key findings \& implications}
\label{tab:Findings}
\begin{tabular}{@{}m{3cm}m{6.8cm}m{6.8cm}@{}}
\toprule
    & \textbf{Findings} & \textbf{Implications} \\
    \midrule

 Working activity distribution WFH compared to a typical office day. & Overall, the ranking among work activities remains unchanged. However, when WFH developers spend less time in: Bugfixing ($t_1=-5.31, t_2=-3.55$), Meetings ($t_1=-9.95, t_2=-6.63$), Breaks ($t_1=-7.39, t_2=-14.30$), Interruptions ($t_2=-5.39$), E-Mails ($t_1=-3.69$), and more time in Specification ($t_1=4.65, t_2=4.05$), ($t_1=4.65, t_2=4.05$), Testing ($t_1=3.41, t_2=3.32$), Administration ($t_1=4.58, t_2=4.28$), Documentation ($t_1=5.18, t_2=5.07$), Learning ($t_1=4.24, t_2=3.38$). & WFH does not affect the time spent on working tasks by software developers and the distribution is comparable to a typical office day. The significant time reduction of meetings suggest that online-meetings are more time-efficient than physical ones. Also, professionals seems to be more focused when working remotely, having fewer interruptions. This allows them, among others, to dedicate more time in developing their own skills.  \\ \addlinespace

    \addlinespace
     Working activity distribution during the pandemic. & Very high correlation of the group averages of time 1 and 2: $r(13) = .99, p < .0001$. Two exceptions were more Breaks ($t=4.71$) and Networking ($t=4.33$) at time 1 compared to time 2.  & Developers had a very regular work activity distribution during the pandemic, which was comparable to their office day. Fewer breaks and networking might depend that professionals adapted to the new situations towards the end of the lockdown, being more time-efficient. \\ \addlinespace

    \addlinespace   

        Effects of well-being and productivity on working activities while working from home.  & Only the relation between Productivity and Breaks was significant at time 1 ($r_1 = -.30, p_1 = .00002$); at time 2 the correlation was again negative but not statistical significant ($r_2 = -.16, p_2 = .03$). & The time spent by software engineers on individual tasks while working from home does not affect their productivity or well-being. In the case of WFH, organizations can plan work activities as they have done so far.  \\ \addlinespace
    
    \addlinespace

    Influence of psychological and social variables on working activities while working remotely. & No significant correlations between any activity and variable ($r < .25, p > .0005$). & We could not associate any indicator of burnout, motivation, disengagement, relation to the organization, and self-organization with the activities performed while working from home. This suggests that WFH is \textit{per se}, not a burden for developers. \\

    \bottomrule

\end{tabular}
\vspace{2em}
\end{table*}

\subsection{Threats to validity}
To conclude this section, we briefly address the most relevant limitations. 

\textit{Reliability}. We investigated our subject matter through a two-wave longitudinal study. Notably, over 90\% of our initial informants also took part in the second wave.
Participants were identified using a multi-stage selection process to ensure (i) they are professionally active software engineers, (ii) data quality, and (iii) that they were working from home during the lockdown.

\textit{Construct validity}. 
To enhance reproducibility, we used the taxonomy by Meyer et al.~\cite{meyer2019today} to define the daily activities of software developers.
Similarly, we used those benchmarks to confront it with working from home setting.
Also, we report the Cronbach's alpha across both waves of ten identified variables, as well as their test-retest reliability.

\textit{Conclusion validity}.
Our conclusions rely on multiple statistical analyses, such as one-sample $t$-tests, paired $t$-tests, and Pearson's correlation.
Furthermore, we also ran a non-parametric Spearman's rank correlations test for our conclusion's consistency since not all distributions were perfectly normally distributed.
To support Open Science, we made the replication package in R and our raw data and openly available on Zenodo.

\textit{Internal validity}.
For this investigation, we used self-reported measures for well-being, productivity, and other psychological and social variables, which might be considered a limitation.
However, similar to Meyer et al., our primary focus was to understand the developers' perspective of what makes ``a good day'' and how individual tasks influence both well-being and productivity while working from home.
The research was performed towards the end of the worldwide lockdown in spring 2020. 
This enabled our participants to report a more mature and stable assessment of the new working setting.
We only considered countries with comparable lockdown measures (e.g., we excluded, among others, Denmark, Germany, and Sweden as these countries did not face a total lockdown or had different measures in place in the country's regions).
Thus, we asked both waves about lockdown conditions in their home country and if they were still working from home.
Since all selected informants faced comparable conditions, we did not exclude any of the 192 selected software professionals.

\textit{External validity}.
We designed this study to maximize internal validity.
Therefore, we determined our sample size with an \textit{a priori} power analysis.
So, we did not work with a representative sample of the software engineering population in mind (such as Russo and Stol~\cite{russo2020gender} did, where the research goal was the generalization of results, surveying over 400 software engineers).
However, we recognize having submitted our surveys in the middle of a very peculiar period. 
This limits the extent to which we can generalize our findings in a non-pandemic working from home setting.
Notwithstanding, we also realize that we require fast and reliable evidence regarding the COVID-19 crisis we are facing right now, improving the quality of developers' daily lives.
This study will also enable a better-informed research design for future remote working studies once this pandemic is over.

\section{Conclusion}
\label{sec:conclusion}

This research focused on software engineers' work distributions during enforced WFH and the association between single tasks with well-being, productivity, and other social and psychological variables.
To do so, we employed a longitudinal study design across two waves. We found that developers still spend proportionally the same amount of time on their different daily activities. 
For example, the software engineers in our sample still spent most of their working time on coding, bugfixing, meetings, testing, and e-mails, as previously reported by Meyer et al.~\cite{meyer2019today}.
Nevertheless, we found some significant mean differences. Our participants reported having spent less time in meetings and breaks, suggesting that both were less common, possibly due to developers' adaption of working remotely.
Similarly, no significant relations have been found between productivity, well-being, and relevant social and psychological variables with working activities.
Based on our findings, our research suggests that WFH does not \textit{per se} presents a challenge for either organizations or developers.

Future research will focus on exploring moderation effects by investigating whether perceived effectiveness, independence, and meaningfulness of a task moderate the relation between each work activity with well-being and productivity and other relevant variables. 
Further, more tailored recommendations based on developers' persona would provide a more nuanced understanding of the subject matter since we only considered average effects in this study.

\section*{Supplementary Materials} 
The full replication package and additional Tables are openly available under CC BY 4.0 license on Zenodo, DOI: \url{https://doi.org/10.5281/zenodo.4104390}.


\section*{Acknowledgment}
This work was supported, in part, by the Carlsberg Foundation under grant agreement number CF20-0322 (PanTra --- Pandemic Transformation).



\bibliographystyle{IEEEtran}
\bibliography{bib}

\begin{thebibliography}{10}
\providecommand{\url}[1]{#1}
\csname url@samestyle\endcsname
\providecommand{\newblock}{\relax}
\providecommand{\bibinfo}[2]{#2}
\providecommand{\BIBentrySTDinterwordspacing}{\spaceskip=0pt\relax}
\providecommand{\BIBentryALTinterwordstretchfactor}{4}
\providecommand{\BIBentryALTinterwordspacing}{\spaceskip=\fontdimen2\font plus
\BIBentryALTinterwordstretchfactor\fontdimen3\font minus
  \fontdimen4\font\relax}
\providecommand{\BIBforeignlanguage}[2]{{%
\expandafter\ifx\csname l@#1\endcsname\relax
\typeout{** WARNING: IEEEtran.bst: No hyphenation pattern has been}%
\typeout{** loaded for the language `#1'. Using the pattern for}%
\typeout{** the default language instead.}%
\else
\language=\csname l@#1\endcsname
\fi
#2}}
\providecommand{\BIBdecl}{\relax}
\BIBdecl

\bibitem{Ralph2020pandemic}
P.~Ralph \emph{et~al.}, ``Pandemic programming: How {COVID}-19 affects software
  developers and how their organizations can help,'' \emph{Empirical Software
  Engineering}, 2020.

\bibitem{ford2020tale}
D.~Ford, M.-A. Storey, T.~Zimmermann, C.~Bird, S.~Jaffe, C.~Maddila, J.~L.
  Butler, B.~Houck, and N.~Nagappan, ``A tale of two cities: Software
  developers working from home during the {COVID-19} pandemic,'' \emph{arXiv
  preprint arXiv:2008.11147}, 2020.

\bibitem{forsgren_2020}
\BIBentryALTinterwordspacing
N.~Forsgren, ``Octoverse spotlight: An analysis of developer productivity, work
  cadence, and collaboration in the early days of {COVID-19} at {GitHub},'' May
  2020. [Online]. Available: \url{tiny.cc/vl5ysz}
\BIBentrySTDinterwordspacing

\bibitem{bao2020does}
L.~Bao, T.~Li, X.~Xia, K.~Zhu, H.~Li, and X.~Yang, ``How does working from home
  affect developer productivity?--a case study of baidu during {COVID-19}
  pandemic,'' \emph{arXiv preprint arXiv:2005.13167}, 2020.

\bibitem{russo2020predictors}
D.~Russo, P.~H.~P. Hanel, S.~Altnickel, and N.~van Berkel, ``Predictors of
  well-being and productivity among software professionals during the
  {COVID-19} pandemic--a longitudinal study,'' \emph{Empirical Software
  Engineering}, 2021.

\bibitem{Walton2020NZadaptation}
S.~Walton, P.~O’Kane, and D.~Ruwhiu, ``{New Zealanders’} attitudes towards
  working from home,'' URL
  \url{https://www.otago.ac.nz/news/news/otago737417.html}, University of
  Otago, Tech. Rep., 2020.

\bibitem{BusinessInsider2020}
\BIBentryALTinterwordspacing
J.~Hadden, L.~Casado, T.~Sonnemaker, and T.~Borden, ``19 {Major} companies that
  have announced employees can work remotely long-term,'' Sep 2020. [Online].
  Available:
  \url{https://www.businessinsider.com/companies-asking-employees-to-work-from-home-due-to-coronavirus-2020}
\BIBentrySTDinterwordspacing

\bibitem{pounder1998homeworking}
C.~Pounder, ``Homeworking: No longer an easy option?'' \emph{Computers \&
  Security}, vol.~17, no.~1, pp. 27--30, 1998.

\bibitem{guo2001special}
H.~Guo, ``Special requirements for software process improvement applied in
  teleworking environments,'' in \emph{Proceedings of the Second Asia-Pacific
  Conference on Quality Software}.\hskip 1em plus 0.5em minus 0.4em\relax IEEE,
  2001, pp. 331--340.

\bibitem{deshpande2016remote}
A.~Deshpande, H.~Sharp, L.~Barroca, and P.~Gregory, ``Remote working and
  collaboration in agile teams,'' in \emph{Proceedings of the International
  Conference on Information Systems}.\hskip 1em plus 0.5em minus 0.4em\relax
  AIS, 2016.

\bibitem{higa2000understanding}
K.~Higa, O.~R.~L. Sheng, B.~Shin, and A.~J. Figueredo, ``Understanding
  relationships among teleworkers'e-mail usage, {E-Mail} richness perceptions,
  and {E-Mail} productivity perceptions under a software engineering
  environment,'' \emph{IEEE Transactions on Engineering Management}, vol.~47,
  no.~2, pp. 163--173, 2000.

\bibitem{herbsleb2007global}
J.~D. Herbsleb, ``Global software engineering: The future of socio-technical
  coordination,'' in \emph{Proceedings of the Future of Software Engineering
  Conference}.\hskip 1em plus 0.5em minus 0.4em\relax IEEE, 2007, pp. 188--198.

\bibitem{vsmite2010empirical}
D.~{\v{S}}mite, C.~Wohlin, T.~Gorschek, and R.~Feldt, ``Empirical evidence in
  global software engineering: a systematic review,'' \emph{Empirical Software
  Engineering}, vol.~15, no.~1, pp. 91--118, 2010.

\bibitem{buffer2020}
\BIBentryALTinterwordspacing
Buffer, ``The 2020 state of remote work,'' 2020. [Online]. Available:
  \url{https://lp.buffer.com/state-of-remote-work-2020}
\BIBentrySTDinterwordspacing

\bibitem{mesaglio_2020}
\BIBentryALTinterwordspacing
M.~Mesaglio, ``4 actions to be a strong leader during {COVID-19} disruption,''
  Mar 2020. [Online]. Available:
  \url{https://www.gartner.com/smarterwithgartner/4-actions-to-be-a-good-leader-during-covid-19-disruption/}
\BIBentrySTDinterwordspacing

\bibitem{meyer2019today}
A.~Meyer, E.~T. Barr, C.~Bird, and T.~Zimmermann, ``Today was a good day: The
  daily life of software developers,'' \emph{IEEE Transactions on Software
  Engineering}, 2019.

\bibitem{Urban2019}
E.~Urban, ``Why most 1:1 meetings are a waste of time, and what to do
  instead,'' \emph{Forbes}, 2019.

\bibitem{bhui2016perceptions}
K.~Bhui, S.~Dinos, M.~Galant-Miecznikowska, B.~de~Jongh, and S.~Stansfeld,
  ``Perceptions of work stress causes and effective interventions in employees
  working in public, private and non-governmental organisations: a qualitative
  study,'' \emph{BJPsych bulletin}, vol.~40, no.~6, pp. 318--325, 2016.

\bibitem{edwards2009value}
J.~R. Edwards and D.~M. Cable, ``The value of value congruence.'' \emph{Journal
  of Applied Psychology}, vol.~94, no.~3, p. 654, 2009.

\bibitem{mccalister2006hardiness}
K.~T. McCalister, C.~L. Dolbier, J.~A. Webster, M.~W. Mallon, and M.~A.
  Steinhardt, ``Hardiness and support at work as predictors of work stress and
  job satisfaction,'' \emph{American Journal of Health Promotion}, vol.~20,
  no.~3, pp. 183--191, 2006.

\bibitem{chyi2018prediction}
T.~Chyi, F.~J.-H. Lu, E.~T. Wang, Y.-W. Hsu, and K.-H. Chang, ``Prediction of
  life stress on athletes’ burnout: the dual role of perceived stress,''
  \emph{PeerJ}, vol.~6, p. e4213, 2018.

\bibitem{stackoverflow_2017}
\BIBentryALTinterwordspacing
A.~Mazzina, ``What it means to be a remote-first company: Stack overflow,'' Feb
  2017. [Online]. Available:
  \url{https://stackoverflow.blog/2017/02/08/means-remote-first-company/}
\BIBentrySTDinterwordspacing

\bibitem{hat_2015}
\BIBentryALTinterwordspacing
RedHat, ``Red hat named a top 100 company for remote jobs by flexjobs,'' Jan
  2015. [Online]. Available:
  \url{https://www.redhat.com/en/blog/red-hat-named-top-100-company-remote-jobs-flexjobs}
\BIBentrySTDinterwordspacing

\bibitem{felstead2017assessing}
A.~Felstead and G.~Henseke, ``Assessing the growth of remote working and its
  consequences for effort, well-being and work-life balance,'' \emph{New
  Technology, Work and Employment}, vol.~32, no.~3, pp. 195--212, 2017.

\bibitem{perez2002benefits}
M.~P. P{\'e}rez, A.~M. S{\'a}nchez, and M.~de~Luis~Carnicer, ``Benefits and
  barriers of telework: perception differences of human resources managers
  according to company's operations strategy,'' \emph{Technovation}, vol.~22,
  no.~12, pp. 775--783, 2002.

\bibitem{ford2019remote}
D.~Ford, R.~Milewicz, and A.~Serebrenik, ``How remote work can foster a more
  inclusive environment for transgender developers,'' in \emph{Proceedings of
  the International Workshop on Gender Equality in Software Engineering}.\hskip
  1em plus 0.5em minus 0.4em\relax IEEE, 2019, pp. 9--12.

\bibitem{james2014secure}
P.~James and D.~Griffiths, ``A secure portable execution environment to support
  teleworking,'' \emph{Information Management \& Computer Security}, 2014.

\bibitem{reynolds_2020}
\BIBentryALTinterwordspacing
B.~W. Reynolds, ``100 top companies with remote jobs in 2015,'' Mar 2020.
  [Online]. Available:
  \url{https://www.flexjobs.com/blog/post/100-top-companies-with-remote-jobs-in-2015/}
\BIBentrySTDinterwordspacing

\bibitem{russo2020gender}
D.~Russo and K.-J. Stol, ``Gender differences in personality traits of software
  engineers,'' \emph{IEEE Transactions on Software Engineering}, vol. In Press,
  p.~16, 2020.

\bibitem{palan2018prolific}
S.~Palan and C.~Schitter, ``Prolific.ac—-a subject pool for online
  experiments,'' \emph{Journal of Behavioral and Experimental Finance},
  vol.~17, pp. 22--27, 2018.

\bibitem{mcdonald2013test}
R.~P. McDonald, \emph{Test theory: A unified treatment}.\hskip 1em plus 0.5em
  minus 0.4em\relax psychology press, 2013.

\bibitem{diener1985satisfaction}
E.~Diener, R.~A. Emmons, R.~J. Larsen, and S.~Griffin, ``The satisfaction with
  life scale,'' \emph{Journal of personality assessment}, vol.~49, no.~1, pp.
  71--75, 1985.

\bibitem{hair2013multivariate}
J.~F. Hair, W.~C. Black, B.~J. Babin, R.~E. Anderson, R.~L. Tatham
  \emph{et~al.}, \emph{Multivariate data analysis}.\hskip 1em plus 0.5em minus
  0.4em\relax Pearson Education, 2013, vol. 7th ed.

\bibitem{wagner2018systematic}
S.~Wagner and M.~Ruhe, ``A systematic review of productivity factors in
  software development,'' \emph{arXiv preprint arXiv:1801.06475}, 2018.

\bibitem{ko2019we}
A.~J. Ko, ``Why we should not measure productivity,'' in \emph{Rethinking
  Productivity in Software Engineering}.\hskip 1em plus 0.5em minus 0.4em\relax
  Springer, 2019, pp. 21--26.

\bibitem{meyer2014software}
A.~N. Meyer, T.~Fritz, G.~C. Murphy, and T.~Zimmermann, ``Software developers'
  perceptions of productivity,'' in \emph{Proceedings of the International
  Symposium on Foundations of Software Engineering}, 2014, pp. 19--29.

\bibitem{Cohen1988perceived}
S.~Cohen, ``Perceived stress in a probability sample of the {U}nited
  {S}tates.'' \emph{The social psychology of health}, 1988.

\bibitem{farmer1986boredom}
R.~Farmer and N.~D. Sundberg, ``Boredom proneness--the development and
  correlates of a new scale,'' \emph{Journal of Personality Assessment},
  vol.~50, no.~1, pp. 4--17, 1986.

\bibitem{struk2017short}
A.~A. Struk, J.~S. Carriere, J.~A. Cheyne, and J.~Danckert, ``A short boredom
  proneness scale: Development and psychometric properties,''
  \emph{Assessment}, vol.~24, no.~3, pp. 346--359, 2017.

\bibitem{ryan2000self}
R.~M. Ryan and E.~L. Deci, ``Self-determination theory and the facilitation of
  intrinsic motivation, social development, and well-being.'' \emph{American
  Psychologist}, vol.~55, no.~1, pp. 68--78, 2000.

\bibitem{sheldon2012balanced}
K.~M. Sheldon and J.~C. Hilpert, ``The balanced measure of psychological needs
  (bmpn) scale: An alternative domain general measure of need satisfaction,''
  \emph{Motivation and Emotion}, vol.~36, no.~4, pp. 439--451, 2012.

\bibitem{hanel2020well}
P.~Hanel, U.~Wolfradt, L.~Wolf, G.~L. D.~H. Coelho, and G.~Maio, ``Well-being
  as a function of person-country fit in human values,'' \emph{Nature
  Communications}, vol.~11, 2020.

\bibitem{hays_statistics_1994}
W.~Hays, \emph{Statistics}, 5th~ed.\hskip 1em plus 0.5em minus 0.4em\relax
  Harcourt Brace, 1994.

\bibitem{gagne2005self}
M.~Gagn{\'e} and E.~L. Deci, ``Self-determination theory and work motivation,''
  \emph{Journal of Organizational behavior}, vol.~26, no.~4, pp. 331--362,
  2005.

\end{thebibliography}

\end{document}